\documentclass[12pt]{article}
\usepackage[english,german,french,polish]{babel}
\usepackage[T1]{fontenc}
\usepackage{amsfonts}

\begin{document}

\selectlanguage{english}

\baselineskip 0.76cm
\topmargin -0.6in
\oddsidemargin -0.1in

\let\ni=\noindent

\renewcommand{\thefootnote}{\fnsymbol{footnote}}

\pagestyle {plain}

\setcounter{page}{1}

\pagestyle{empty}

~~~

\begin{flushright}
IFT-04/32
\end{flushright}

{\large\centerline{\bf Embedding the neutrino four-group $Z_2\times Z_2$ }}

{\large\centerline{\bf into the alternating group of four objects $A_4${\footnote {Work supported in part by the Polish State Committee for Scientific Research (KBN), grant 2 P03B 129 24 (2003--2004).}}}}

\vspace{0.4cm}

{\centerline {\sc Wojciech Kr\'{o}likowski}}

\vspace{0.3cm}

{\centerline {\it Institute of Theoretical Physics, Warsaw University}}

{\centerline {\it Ho\.{z}a 69,~~PL--00--681 Warszawa, ~Poland}}

\vspace{1.0cm}

{\centerline{\bf Abstract}}

\vspace{0.2cm}

If and only if the small neutrino mixing sine $s_{13}$ is put zero, the effective neutrino mass matrix
$M$ with any mixing sines $s_{12}$ and $s_{23}$ is invariant under a cyclic group $Z_2$ of the order two and --- in the limit of $m_2 - m_1 \rightarrow 0$ --- also under the group $Z_2\times Z_2$ 
often called the four-group. However, the elements of this group do not build up fully the term of $M$ 
proportional to $m_2 - m_1$. When the four-group is embedded into the group of even permutations of 
four objects $A_4$ (which is of the order twelve), then the term of $M$ proportional to $m_2 - m_1$ can be fully constructed from elements of $A_4$. In contrast, when the four-group is embedded into the 
dihedral group $D_4$ of the order eight, then the term of $M$ proportional to $m_2 - m_1$ cannot be 
fully built up from elements of $D_4$.

\vspace{0.5cm}

\ni PACS numbers: 12.15.Ff , 14.60.Pq , 12.15.Hh .

\vspace{0.8cm}

\ni December 2004 

\vfill\eject

~~~
\pagestyle {plain}

\setcounter{page}{1}

\vspace{0.3cm}

\ni {\bf 1. Introduction}

\vspace{0.3cm}

As is well known, the generic form of effective neutrino mixing matrix $U = \left( U_{\alpha i} \right) \;(\alpha = e, \mu, \tau\;,\; i=1, 2, 3)$ reads, 

%rownanie 1*
\begin{equation}
U = \left( \begin{array}{ccc} c_{12} & s_{12} & 0 \\ - c_{23} s_{12} & c_{23} c_{12} & s_{23} \\ s_{23} s_{12} & -s_{23} c_{12} & c_{23}  \end{array} \right) \;,
\end{equation}

\vspace{0.2cm}

\ni if $s_{13}$ is put zero consistently with the nonobservation of the neutrino oscillations in the 
short-baseline experiments for reactor $\bar{\nu}_e$'s (the estimated upper bound is $s^2_{13} < 0.04$) [1]. Here, $c_{ij} = \cos \theta_{ij}$ and $s_{ij} = \sin \theta_{ij}$. Two possible CP-violating Majorana phases irrelevant for the oscillation processes are omitted in Eq. (1). One CP-violating Dirac phase relevant for these processes disappears together with $s_{13} = 0$.

The formula (1) is consistent with all well established neutrino-oscillation experiments that provide the global estimates: $\theta_{12} \sim 32^\circ $ and $\theta_{23} \sim 45^\circ $ for mixing angles, and $\Delta m^2_{21} = m^2_2 - m^2_1 \sim 8\times 10^{-5}\;{\rm eV}^2$  and $|\Delta m^2_{32}| = |m^2_3 - m^2_2| \sim 2.5\times 10^{-3}\;{\rm eV}^2$ for mass-squared differences [2]. 

The unitary matrix $U$ gives the neutrino mixing $\nu_\alpha  = \sum_i U_{\alpha i}\, \nu_i$, where $ \nu_\alpha = \nu_e, \nu_\mu, \nu_\tau$ and $\nu_i=\nu_1, \nu_2, \nu_3$ are the flavor and mass active neutrinos, respectively. In the flavor representation, where the charged-lepton mass matrix is diagonal, the matrix $U$ gives also the diagonalization $ \sum_{\alpha, \beta} U^*_{\alpha i} M_{\alpha\,\beta}  U_{\beta j} = m_i \delta_{ij}$ of the effective neutrino mass matrix  $ M = \left(M_{\alpha \beta}\right)\;(\alpha, \beta = e, \mu, \tau)$. Then, the inverse formula $M_{\alpha \beta}  = \sum_i U_{\alpha i}\, m_i\, U^*_{\beta i}$ enables one to calculate the mass-matrix elements $M_{\alpha \beta}$ in terms of neutrino masses $m_i$ and mixing-matrix elements $U_{\alpha i}$. In this way, using the form (1) of $U$, one obtains the following generic form of effective neutrino mass-%
matrix valid in the case of $s_{13} = 0$:

%rownanie 2*
\begin{eqnarray}
M & = & \frac{m_1+m_2}{2} \left( \begin{array}{rrr} 1 & 0 & 0 \\ 0 & c^2_{23} & -c_{23} s_{23} \\ 0 & -c_{23} s_{23} & s^2_{23} \end{array} \right) + m_3 \left( \begin{array}{rrr} 0 & 0 & 0 \\ 0 & s^2_{23} & c_{23} s_{23}  \\ 0 & c_{23} s_{23}  & c^2_{23} \end{array}\right) \nonumber \\ & & \nonumber \\ & &  \!\!\!+ \frac{m_2-m_1}{2} \left[c_{\rm sol} \left( \begin{array}{rrr} -1 & 0 & 0 \\ 0 & c^2_{23} & -c_{23} s_{23} \\ 0 & -c_{23} s_{23} & s^2_{23} \end{array} \right) + s_{\rm sol} \left( \begin{array}{rrr} 0 & c_{23} & -s_{23} \\ c_{23} & 0 & 0  \\ -s_{23} & 0 & 0 \end{array}\right) \right] ,
\end{eqnarray}

\ni where $c_{\rm sol} = c^2_{12} - s^2_{12} = \cos 2\theta_{12} $ and $s_{\rm sol} = 2c_{12} s_{12} = \sin 2\theta_{12}$.

In the formula (2), all three $3\times 3$ matrices standing at {\small $\frac{1}{2}$}$(m_1+m_2)$, $m_3$ and {\small $\frac{1}{2}$}$(m_2-m_1)$ commute, while two components of the third $3\times 3$ matrix anticommute. Diagonalizing both sides of Eq. (2), one gets consistently

%rownanie 3*
\begin{eqnarray}
\left( \begin{array}{rrr} m_1 & 0 & 0 \\ 0 & m_2 & 0 \\ 0 & 0 & m_3 \end{array} \right) & = & \frac{m_1 + m_2}{2} \left( \begin{array}{rrr} 1 & 0 & 0 \\ 0 & 1 & 0 \\ 0 & 0 & 0 \end{array} \right) + \,m_3 \left( \begin{array}{rrr} 0 & 0 & 0 \\ 0 & 0 & 0 \\ 0 & 0 & 1 \end{array} \right) \nonumber \\ &+ & \frac{m_2 - m_1}{2} \left( \begin{array}{rrr} \!\!-1 & 0 & 0 \\ \!\!0 & 1 & 0 \\ \!\!0 & 0 & 0 \end{array} \right) \,, 
\end{eqnarray}

\ni where $ U^\dagger M U =$ diag$(m_1, m_2, m_3)$.

\vspace{0.3cm}

\ni {\bf 2. The four-group $Z_2\times Z_2$}

\vspace{0.3cm}

The effective neutrino mass matrix (2) valid in the case of $s_{13} = 0$ can be rewritten as follows:

\vspace{-0.2cm}

%rownanie 4*
\begin{eqnarray}
M & = & \;\,\frac{m_1+m_2}{2}\frac{1}{2}\left(\varphi_1 - \varphi_4\right) + m_3 \frac{1}{2}\left(\varphi_1 + \varphi_4\right)  \nonumber \\ 
& & + \frac{m_2-m_1}{2} \frac{1}{2} \left[ c_{\rm sol} \left( \varphi_2 - \varphi_3 \right) + s_{\rm sol}\left(c_{23}\lambda_1 - s_{23}\lambda_4 \right)\right]\,,
\end{eqnarray}

\ni if the $3\times 3$ matrices

%rownanie 5*
\begin{eqnarray} 
\varphi_1 = \left( \begin{array}{rrr} \; 1 & 0 & 0 \\ 0 & \,\;\;1 & 0  \\ 0 & 0 & \,\;\;1 \end{array} \right) & , &
\varphi_2 = \left( \begin{array}{rrr} -1 & 0 & 0 \\ 0 & c_{\rm atm} & -s_{\rm atm} \\ 0 & -s_{\rm atm} & -c_{\rm atm} \end{array} \right) \;\,, \nonumber \\
\varphi_3 = \left( \begin{array}{rrr} 1 & 0 & 0 \\ 0 & -1 & 0 \\ 0 & 0 & -1 \end{array} \right) & , &
\varphi_4 = \left( \begin{array}{rrr} -1 & 0 & 0 \\ 0 & -c_{\rm atm} & \;\;s_{\rm atm} \\ 0 & s_{\rm atm} & c_{\rm atm} \end{array} \right) 
\end{eqnarray}

\vspace{0.2cm}

\ni and

%rownanie 6
\begin{equation}
\lambda_1 = \left( \begin{array}{rrr} 0 & 1 & 0 \\ 1 & 0 & 0 \\ 0 & 0 & 0 \end{array} \right) \;,\;
\lambda_4 = \left( \begin{array}{rrr}  0 & 0 & 1 \\ 0 & 0 & 0 \\ 1 & 0 & 0 \end{array} \right) 
\end{equation}

\vspace{0.2cm}

\ni are introduced, where $c_{\rm atm} = c^2_{23} - s^2_{23} = \cos 2\theta_{23}$ and $s_{\rm atm} = 2c_{23} s_{23} = \sin 2\theta_{23}$. Here, the values $c_ {23} = 1/\sqrt2 = s_{23}$ ({\it i.e.}, $c_{\rm atm} =0$ and $s_{\rm atm} = 1$) are consistent with experimental estimates.

It is easy to check for any $c_{\rm atm} $ and $s_{\rm atm}$ that four $3\times 3$ matrices (5) are mutually multiplied according to the Cayley table of the form 

\vspace{0.2cm}

\begin{center}
\begin{tabular}{l|llll}
 & $\varphi_1$ & $\varphi_2$ & $\varphi_3$ & $\varphi_4$ \\ & & & & \\ \hline
$\varphi_1$ & $\varphi_1$ & $\varphi_2$ & $\varphi_3$ & $\varphi_4$ \\
$\varphi_2$ & $\varphi_2$ & $\varphi_1$ & $\varphi_4$ & $\varphi_3$ \\
$\varphi_3$ & $\varphi_3$ & $\varphi_4$ & $\varphi_1$ & $\varphi_2$ \\ 
$\varphi_4$ & $\varphi_4$ & $\varphi_3$ & $\varphi_2$ & $\varphi_1$ \\
\end{tabular}
\end{center}

\vspace{0.2cm}

\ni what characterizes the Abelian finite group $Z_2\times Z_2$ of the order four, often called the four-group [3]. It is isomorphic to the simplest dihedral group of the order four. Thus, for any $c_{\rm atm} $ and $s_{\rm atm}$, four $3\times 3$ matrices (5) constitute a reducible representation $\underline{3}$ of the Abelian four-group [4]. In fact, they can be simultaneously diagonalized, leading to a reduced representation $\underline{3} = \underline{1} + \underline{1} + \underline{1} =$ diag($\underline{1}, \underline{1},\underline{1}$).

One can see from Eq. (4) that the term of $M$ proportional to $m_2 - m_1$ is not fully built up from elements of four-group: the $3\times 3$ matrix $c_{23}\lambda_1 - s_{23}\lambda_4 $ lies outside the four-group. This matrix commutes with $\varphi_1$ and $\varphi_4$, and anticommutes with $\varphi_2$ and $\varphi_3$.

It is not difficult to verify for any $c_{\rm atm} $ and $s_{\rm atm}$ that the formula (4) implies four discrete invariances of the effective neutrino mass matrix $M$, two exact:

%rownanie 7*
\begin{equation}
\varphi_{1,4} M \varphi_{1,4} = M 
\end{equation}

\ni  and two valid in the limit of $m_2 - m_1 \rightarrow 0$:

%rownanie 8*
\begin{equation}
\varphi_{2,3} M \varphi_{2,3} = M - (m_2 - m_1) s_{\rm sol} (c_{23} \lambda_1 - s_{23} \lambda_4)
\stackrel{m_2-m_1 \rightarrow 0}{\!\longrightarrow} M 
\end{equation}

\vspace{0.2cm}

\ni (of course, in the case of $\varphi_1$ the invariance is trivial). Thus, for any $c_{\rm atm} $ and $s_{\rm atm}$, the four-group is the symmetry group of $M$ in the limit of $m_2 - m_1 \rightarrow 0$.

Further on, in consistency with experimental estimates, one will put $c_{23} =1/\sqrt2 = s_{23}$ ({\it i.e.}, $c_{\rm atm} = 0$ and $s_{\rm atm} = 1$).

\vfill\eject

\vspace{0.3cm}

\ni {\bf 3. The alternating group of four objects $A_4$}

\vspace{0.2cm}

The group of even permutations of four objects $A_4$ was applied to the neutrino mass matrix in Refs. [5]. This non-Abelian finite group consists of twelve permutations

%rownanie 9*
\begin{eqnarray} 
(1,2,3,4) \;\;,\;\;  (2,1,4,3) & , &  (3,4,1,2) \;\;,\;\;  (4,3,2,1)  \;\;, \nonumber \\
(2,3,1,4) \;,\;\;\;  (1,4,2,3) & , &  (4,1,3,2) \;\;,\;\;  (3,2,4,1)  \;\;, \nonumber \\
(3,1,2,4) \;\;,\;\;  (4,2,1,3) & , &  (1,3,4,2) \;\;,\;\;  (2,4,3,1) 
\end{eqnarray}

\ni of which the first four constitute an Abelian subgroup of $A_4$, isomorphic to the four-group $Z_2\times Z_2$. The twelve permutations (9) can be represented, respectively, by the following twelve $4\times 4$ matrices constructed from the formal Dirac matrices in the Dirac representation:

%rownanie 10*
\begin{eqnarray} 
g_1 = {\bf 1}^{(D)} \;\;,\;\; g_2 = \sigma_1^{(D)} & , & g_3 = \gamma_5 \;\;,\;\; g_4 = \gamma_5 \sigma_1^{(D)}  \;\;, \nonumber \\
g_5 =\left( \frac{{\bf 1}^{(D)}-\beta}{2}+\frac{{\bf 1}^{(D)}+\beta}{2}\sigma_1^{(D)} \right) \!\!\!\!\!\! & & \!\!\!\! \!\! \left( \frac{{\bf 1}^{(D)}-\sigma_3^{(D)}}{2}+\gamma_5 \frac{{\bf 1}^{(D)}+\sigma_3^{(D)}}{2}\right) \;\;, \nonumber \\
 g_6 =  \gamma_5 g_5 \;\;,\;\; g_7 & \!\!=\!\! &  g_5 \gamma_5 \;\;,\;\; g_8 = \sigma_1^{(D)} g_5 \;\;, \nonumber \\ 
g_9 = \frac{{\bf 1}^{(D)}-\beta}{2} \left({\bf 1}^{(D)}+\gamma_5 \sigma_1^{(D)} \right)\frac{{\bf 1}^{(D)}-\sigma_3^{(D)}}{2} & \!\!+\!\! & \frac{{\bf 1}^{(D)}+\beta}{2} \left({\bf 1}^{(D)}+\gamma_5 \sigma_1^{(D)} \right)\frac{{\bf 1}^{(D)}+\sigma_3^{(D)}}{2} \;\;, \nonumber \\
 g_{10} = g_9 \sigma_1^{(D)} \;\;,\;\; g_{11} & \!\!=\!\! &  g_9 \gamma_5 \;\;,\;\; g_{12} = \gamma_5 g_9 \;,
\end{eqnarray}

\ni where ${\bf 1}^{(D)} =$ diag(${\bf 1}^{(2)},{\bf 1}^{(2)}$) , $\gamma_5 =$ antidiag(${\bf 1}^{(2)},{\bf 1}^{(2)}$) , $\beta = $ diag(${\bf 1}^{(2)},-{\bf 1}^{(2)}$) and $\sigma_{1,3}^{(D)} =$ diag($\sigma_{1,3},\sigma_{1,3}$) with ${\bf 1}^{(2)} =$ diag(1,1) , $\sigma_1 =$ antidiag(1,1) and $\sigma_3 =$ diag(1,-1).

Twelve $4\times 4$ matrices (10) form a reducible representation $\underline{4}$ of the non-Abelian group $A_4$. In fact, it can be reduced to a representation $\underline{4} = \underline{1} + \underline{3} = $ diag($\underline{1},\underline{3}$) through the unitary transformation realizing the transitions

%rownanie 11*
\begin{equation}
\gamma_5 \rightarrow {\rm diag}({\bf 1}^{(2)},-{\bf 1}^{(2)}) \;,\; \beta \rightarrow {\rm antidiag}(-{\bf 1}^{(2)},-{\bf 1}^{(2)})
\end{equation}

\ni and 

%rownanie 12*
\begin{equation}
\sigma_1^{(D)} \rightarrow {\rm diag}(\sigma_3,\sigma_3) \;,\; \sigma_3^{(D)} \rightarrow {\rm diag}(-\sigma_1,-\sigma_1)
\end{equation}

\ni Then $g_a \rightarrow g'_a\;(a = 1,2,...,12)$ with $g'_a =$ diag($1,\varphi'_a$), where $\varphi'_a\;(a = 1,2,...,12)$ are the following twelve $3\times 3$ matrices:

%rownanie 13*
$$
\begin{array} {rrrr}
\varphi'_1 = \left(\!\begin{array}{rrr}1 & 0 & 0 \\ 0 & 1 & 0 \\ 0 & 0 & 1 \end{array}\!\!\right), & \!\!\varphi'_2 = \left(\!\begin{array}{rrr}\!\!-1 & 0 & 0  \\ 0 & 1 & 0 \\ 0 & 0 &\!\!-1 \end{array}\!\!\right), & \!\!\varphi'_3 = \left(\!\begin{array}{rrr}1 & 0 & 0 \\  0 &\!\!-1 & 0 \\ 0 & 0 &\!\!-1 \end{array}\!\!\right), & \!\!\varphi'_4 = \left(\!\begin{array}{rrr}\!\!-1 & 0 & 0 \\ 0 &\!\!-1 & 0 \\ 0 & 0 & 1 \end{array}\!\!\right), \\ & & & \\
\varphi'_5 = \left(\!\begin{array}{rrr}0 & 1 & 0  \\ 0 & 0 & 1 \\ 1 & 0 & 0  \end{array}\!\!\right), & \!\!\varphi'_6 = \left(\!\begin{array}{rrr} 0 & 1 & 0  \\ 0 & 0 &\!\!-1 \\ \!\!-1 & 0 & 0 \end{array}\!\!\right), & \!\!\varphi'_7 = \left(\!\begin{array}{rrr} 0 &\!\!-1 & 0  \\ 0 & 0 &\!\!-1 \\ 1 & 0 & 0 \end{array}\!\!\right), & \!\!\varphi'_8 = \left(\!\begin{array}{rrr} 0 &\!\!-1 & 0  \\ 0 & 0 & 1 \\ \!\!-1 & 0 & 0 \end{array}\!\!\right), \\ & & & \\
\varphi'_9 = \left(\!\begin{array}{rrr} 0 & 0 & 1 \\ 1 & 0 & 0 \\ 0 & 1 & 0 \end{array}\!\!\right), &  \!\!\varphi_{10} = \left(\!\begin{array}{rrr} 0 & 0 &\!\!-1 \\ \!\!-1 & 0 & 0 \\ 0 & 1 & 0 \end{array}\!\!\right), &   \!\!\varphi'_{11} = \left(\!\begin{array}{rrr} 0 & 0 &\!\!-1 \\ 1 & 0 & 0 \\ 0 &\!\!-1 & 0 \end{array}\!\!\right), & \!\!\varphi'_{12} = \left(\!\begin{array}{rrr} 0 & 0 & 1 \\ \!\!-1 & 0 & 0 \\ 0 &\!\!-1 & 0 \end{array}\!\!\right), 
\end{array}\eqno(13)
$$

\addtocounter{equation}{1}

\ni forming an irreducible representation $\underline{3}$ of the group $A_4$. Its irreducible representation $\underline{1}$ involved in $g'_a \;(a=1,2,...,12)$ consists of twelve numbers 1,1,...,1. There are also two other irreducible representations $\underline{1}$ of $A_4$, not involved in the $g'_a =$ diag($1,\varphi'_a$) considered here.

Note the identities

\vspace{-0.2cm}

%rownanie 14*
\begin{equation}
{\varphi'}_5^T = {\varphi'}_9 \;,\; {\varphi'}_6^T = {\varphi'}_{11} \;,\; {\varphi'}_7^T = {\varphi'}_{12} \;,\; {\varphi'}_8^T = {\varphi'}_{10} 
\end{equation}

\ni and

\vspace{-0.2cm}

%rownanie 15*
\begin{eqnarray} 
\varphi'_1 + \varphi'_2 + \varphi'_3 + \varphi'_4 & = & 0 \;\;, \nonumber \\
\varphi'_5 + \varphi'_6 + \varphi'_7 + \varphi'_8 & = & 0 \;\;, \nonumber \\
\varphi'_9 + \varphi'_{10} + \varphi'_{11} + \varphi'_{12} & = & 0 \;\;, 
\end{eqnarray}

\ni as well as

\vspace{-0.2cm}

%rownanie 16*
\begin{equation}
\frac{1}{2}\left(\varphi'_5 + \varphi'_6 + \varphi'_9 + \varphi'_{11} \right) = \left(\begin{array}{rrr} 0 & 1 & 0 \\ 1 & 0 & 0 \\ 0 & 0 & 0 \end{array}\right) = \lambda_1.
\end{equation}

\vspace{0.2cm}

\ni The matrix $\lambda_1$ commutes with $\varphi'_1$ and $\varphi'_4$, and anticommutes with $\varphi'_2$ and $\varphi'_3$. 

From Eqs. (15) one gets the "strong"\, constraint

%rownanie 17*
\begin{equation}
\sum_a g'_a = 12\left(\begin{array}{rrrr} 1 & 0 & 0 & 0 \\ 0 & 0 & 0 & 0 \\ 0 & 0 & 0 & 0 \\ 0 & 0 & 0 & 0 \end{array}\right) \,.
\end{equation}

\ni Thus, the "weak"\, constraint

%rownanie 18*
\begin{equation}
\sum_a g'_a \left(\begin{array}{r} 1 \\ 2 \\ 3 \\ 4  \end{array}\right) = 0 \,,
\end{equation}

\ni if imposed on the state $(1,2,3,4)^T$ of four objects, implies that the state $(1,0,0,0)^T$ of the object 1 vanishes, while the state $(0,2,3,4)^T$ of the objects 2,3,4 is not constrained. When interpreting 1 as a light sterile neutrino $\nu_s$ and 2,3,4 as three active neutrinos $\nu_e, \nu_\mu, \nu_\tau $, the "weak"\, constraint (18) may eliminate $\nu_s$ as an "unphysical"\, object, leaving only $\nu_e, \nu_\mu, \nu_\tau $ as "physical". However, in absence of the "weak"\, constraint (18) all four neutrinos may be "physical"\, objects.

Another irreducible representation $\underline{3}$ of $A_4$, isomorphic to the previous $\underline{3} $ consisting of twelve $3\times 3$ matrices (13), can be defined through the unitary transformation

%rownanie 19*
\begin{equation}
\varphi_a =  \left(\begin{array}{rrr}  1 & 0 & 0  \\ 0 & \frac{1}{\sqrt2} & \frac{1}{\sqrt2} \\ 0 & -\frac{1}{\sqrt2} & \frac{1}{\sqrt2}   \end{array}\right) \varphi'_a \left(\begin{array}{rrr}  1 & 0 & 0  \\ 0 & \frac{1}{\sqrt2} & -\frac{1}{\sqrt2} \\ 0 & \frac{1}{\sqrt2} & \frac{1}{\sqrt2} \end{array}\right) \;\;(a = 1,2,...,12)\;.
\end{equation}

\ni Then, one obtains twelve new $3\times 3$ matrices also forming an irreducible representation $\underline{3}$ of $A_4$:

%rownanie 20*
\begin{eqnarray} 
\varphi_1 = \left(\!\begin{array}{rrr}1 & 0 & 0 \\ 0 & 1 & 0 \\ 0 & 0 & 1 \end{array}\!\right) \;,\; 
\varphi_2 = \left(\!\begin{array}{rrr} \!\!-1 & 0 & 0 \\ 0 & 0 &\!\!-1 \\ 0 & \!\!-1 & 0  \end{array}\!\right)\!\! & , & \!\!\varphi_3 = \left(\!\begin{array}{rrr}1 & 0 & 0 \\  0 & \!\!-1 & 0 \\ 0 & 0 & \!\!-1 \end{array}\!\right) \;,\; \varphi_4 = \left(\!\begin{array}{rrr}\!\!-1 & 0 & 0 \\ 0 & 0 & 1 \\ 0 &1 & 0 \end{array}\!\right), \nonumber \\
& & \nonumber \\
\varphi_{5,8} = \frac{1}{\sqrt2} \left(\!\begin{array}{rrr} 0 & \pm 1 & \mp 1  \\ \pm 1 & \frac{1}{\sqrt2} & \frac{1}{\sqrt2} \\ \pm 1 & -\frac{1}{\sqrt2} & -\frac{1}{\sqrt2}  \end{array}\!\right)\!\! & , & \!\!\varphi_{6,7} = \frac{1}{\sqrt2}\left(\!\begin{array}{rrr} 0 & \pm 1 & \mp 1  \\ \mp 1 & -\frac{1}{\sqrt2} & -\frac{1}{\sqrt2} \\ \mp 1 & \frac{1}{\sqrt2} & \frac{1}{\sqrt2} \end{array}\!\right) \;\,, \nonumber \\ 
& & \nonumber \\
\varphi_{9,10} = \frac{1}{\sqrt2}\left(\!\begin{array}{rrr} 0 & \pm 1 & \pm 1 \\ \pm 1 & \frac{1}{\sqrt2} & -\frac{1}{\sqrt2} \\ \mp 1 & \frac{1}{\sqrt2} & -\frac{1}{\sqrt2} \end{array}\!\right)\!\! & , & \!\!\varphi_{11,12} = \frac{1}{\sqrt2}\left(\!\begin{array}{rrr} 0 & \mp 1 & \mp 1 \\ \pm 1 & -\frac{1}{\sqrt2} & \frac{1}{\sqrt2} \\ \mp 1 & -\frac{1}{\sqrt2} & \frac{1}{\sqrt2} \end{array}\!\right) \;\,.
\end{eqnarray}

\vspace{0.2cm}

Notice the identities

%rownanie 21*
\begin{equation}
{\varphi}_5^T = {\varphi}_9 \;,\; {\varphi}_6^T = {\varphi}_{11} \;,\; {\varphi}_7^T = {\varphi}_{12} \;,\; {\varphi}_8^T = {\varphi}_{10} 
\end{equation}

\ni and

%rownanie 22*
\begin{eqnarray} 
\varphi_1 + \varphi_2 + \varphi_3 + \varphi_4 & = & 0 \;\;, \nonumber \\
\varphi_5 + \varphi_6 + \varphi_7 + \varphi_8 & = & 0 \;\;, \nonumber \\
\varphi_9 + \varphi_{10} + \varphi_{11} + \varphi_{12} & = & 0 \;\;, 
\end{eqnarray}

\ni as well as

%rownanie 23*
\begin{eqnarray}
\frac{1}{2}\left(\varphi_1 - \varphi_4 \right) & = & \left(\begin{array}{rrr} 1 & 0 & 0 \\ 0 & \frac{1}{2} & -\frac{1}{2} \\ 0 & -\frac{1}{2} & \frac{1}{2} \end{array}\right) \;,\; \frac{1}{2}\left(\varphi_1 + \varphi_4 \right) = \left(\begin{array}{rrr} 0 & 0 & 0 \\ 0 & \frac{1}{2} & \frac{1}{2} \\ 0 & \frac{1}{2} & \frac{1}{2} \end{array}\right) \,, \nonumber \\ \nonumber \\ & & \frac{1}{2}\left(\varphi_2 - \varphi_3 \right) =  \left(\begin{array}{rrr} -1 & 0 & 0 \\ 0 & \frac{1}{2} & -\frac{1}{2} \\ 0 & -\frac{1}{2} & \frac{1}{2} \end{array}\right) 
\end{eqnarray}

\ni and

%rownanie 24
\begin{equation}
\frac{1}{2}\left(\varphi_5 + \varphi_6 + \varphi_9 + \varphi_{11} \right) = \frac{1}{\sqrt2}\left(\begin{array}{rrr} 0 & 1 & -1 \\ 1 & 0 & 0 \\ -1 & 0 & 0 \end{array}\right) = \frac{1}{\sqrt2}(\lambda_1 - \lambda_4).
\end{equation}

\ni The matrix $(\lambda_1 - \lambda_4)/\sqrt2 $ commutes with $\varphi_1$ and $\varphi_4$, and anticommutes with $\varphi_2$ and $\varphi_3$.

In the case of $c_{23} =1/\sqrt2 = s_{23}$ ({\it i.e.}, $c_{\rm atm} = 0$ and $s_{\rm atm} = 1$), the effective neutrino mass matrix (2) takes the form

%rownanie 25*
\begin{eqnarray}
M & = & \frac{m_1 + m_2}{2} \left( \begin{array}{rrr} 1 & 0 & 0 \\ 0 & \frac{1}{2} & \frac{1}{2} \\ 0 & -\frac{1}{2} & \frac{1}{2} \end{array} \right) + \,m_3 \left( \begin{array}{rrr} 0 & 0 & 0 \\ 0 & \frac{1}{2} & \frac{1}{2} \\ 0 & \frac{1}{2} & \frac{1}{2} \end{array} \right) \nonumber \\ 
&+ & \frac{m_2 - m_1}{2} \left[c_{\rm sol} \left( \begin{array}{rrr} \!\!-1 & 0 & 0 \\ \!\!0 & \frac{1}{2} & -\frac{1}{2} \\ \!\!0 & -\frac{1}{2} & \frac{1}{2} \end{array} \right) + s_{\rm sol} \frac{1}{\sqrt2}\left( \begin{array}{rrr} 0 & 1 & \!\!-1 \\ 1 & 0 & 0 \\ \!\!-1 & 0 & 0 \end{array} \right)\right] \,.
\end{eqnarray}

\ni Thus, making use of Eqs. (23) and (24), one can write

%rownanie 26*
\begin{eqnarray}
M & = & \;\,\frac{m_1+m_2}{2}\frac{1}{2}\left(\varphi_1 - \varphi_4\right) + m_3 \frac{1}{2}\left(\varphi_1 + \varphi_4\right)  \nonumber \\ 
& & + \frac{m_2-m_1}{2} \frac{1}{2} \left[ c_{\rm sol} \left( \varphi_2 - \varphi_3 \right) + s_{\rm sol}\left(\varphi_5 + \varphi_6 + \varphi_9 + \varphi_{11} \right)\right]\,.
\end{eqnarray}

\ni Here, the identities

%rownanie 27*
\begin{equation}
\varphi_1 + \varphi_4 = -(\varphi_2 + \varphi_3) \;,\; \varphi_5 + \varphi_6 + \varphi_9 + \varphi_{11} = 
-\left(\varphi_7 + \varphi_8 + \varphi_{10} + \varphi_{12} \right)
\end{equation}

\ni hold due to Eqs. (22).

In the formula (26) there appear at {\small $\frac{1}{2}$}$(m_1+m_2)$, $m_3$ and {\small $\frac{1}{2}$}$(m_2-m_1)$ three commuting combinations of eight (from twelve) $3\times 3$ matrices $\varphi_a$ given in Eqs. (20). In the third combination there are two anticommuting components since $\varphi_2 - \varphi_3$ and $\varphi_5 + \varphi_6 + \varphi_9 + \varphi_{11} = (\lambda_1 - \lambda_4) \sqrt2 $ anticommute. For the effective neutrino mass matrix $M$ all these combinations play the role of dynamical variables.

In conclusion, one can see from Eq. (26) that the term of $M$ proportional to $m_2 - m_1$ is fully constructed from elements of the group $A_4$ embedding the four-group: the $3\times 3$ matrix $\lambda_1 - \lambda_4 $ lies inside the group $A_4$. As can be inferred from an analogical discussion for the dihedral group $D_4$ of the order eight, also embedding the four-group, the term of $M$ proportional to $m_2 - m_1$ is not fully built up from elements of the group $D_4$: the $3\times 3$ matrix $\lambda_1 - \lambda_4 $ lies outside the group $D_4$ ({\it cf.} Appendix in Ref. [4]). The group $D_4$ was applied to the neutrino mass matrix in Refs. [6].

Of course, in the simple case of four-group as well as in the more involved cases of group $A_4$ and group $D_4$, the matrix $\lambda_1 - \lambda_4 $ appearing in the term of $M$ proportional to $m_2 - m_1$ violates the invariance of $M$ under the transformations generated by matrices $\varphi_2$ and $\varphi_3$, unless the limit of $m_2 - m_1 \rightarrow 0$ is considered ({\it cf.} Eq. (8) with $c_{23} = 1/\sqrt2 = s_{23}$).

As is well known, the neutrino mass matrix, being nonzero, breaks the electroweak gauge symmetry $SU(2)\times U(1)$ as do the mass matrices of charged leptons as well as up and down quarks. Thus, it is natural to expect that, in general, the effective family discrete symmetries of fermion mass matrices --- appearing after the electroweak symmetry is broken --- do not preserve the electroweak symmetry. In particular, the effective family symmetry $Z_2\times Z_2$ of neutrino mass matrix works: under $\varphi_2$ and $\varphi_3$ transformations in the limit of $m_2 = m_1$, and always under $\varphi_1$ and $\varphi_4$ transformations. But, at the same time, in the flavor representation, the charged-lepton mass matrix is diagonal with the entries $m_e \ll m_\mu \ll m_\tau $ excluding its potential invariance under $\varphi_2$ and $\varphi_4$ transformations valid only at $m_\tau = m_\mu$ (for any $c_{\rm atm} $ and $s_{\rm atm}$), though it is always invariant under diagonal $\varphi_1$ and $\varphi_3$ transformations.

\vfill\eject

~~~~

\vspace{0.4cm}

{\centerline{\bf References}}

\vspace{0.4cm}

{\everypar={\hangindent=0.7truecm}
\parindent=0pt\frenchspacing

{\everypar={\hangindent=0.7truecm}
\parindent=0pt\frenchspacing

[1]~M. Apollonio {\it et al.} (Chooz Collaboration), {\it Eur. Phys. J.} {\bf C 27}, 331 (2003).

\vspace{2.0mm}

[2] For a recent review {\it cf.} M. Maltoni {\it et al.}, {\tt hep--ph/0405172}; M.C. Gonzalez-Garcia, {\tt hep--ph/0410030}; G. Altarelli, {\tt hep--ph/0410101}.

\vspace{2.0mm}

[3]~E.P. Wigner, {\it Group theory and its application to the quantum mechanics of atomic spectra}, Academic Press, New York and London, 1959; B.~Simon, {\it Representations of finite and compact groups}, American Mathematical Society,1996.

\vspace{2.0mm}

[4]~W. Kr\'{o}likowski, {\tt hep--ph/0410257} (to appear in {\it Acta Phys. Pol.} {\bf B }, where Appendix is added); and references therein.

\vspace{2.0mm}

[5]~E. Ma, {\tt hep--ph/0307016}; {\tt hep--ph/0409075}; and references therein.

\vspace{2.0mm}

[6]~W. Grimus, L. Lavoura, {\it Acta Phys. Pol.}, {\bf B 34}, 5393 (2003); W. Grimus, A.S.~Yoshi\-pura, 
S.~Kaneko, L. Lavoura, M. Tanimoto, {\tt hep--ph/0407112}; W. Grimus, A.S. Yoshipura, S.~Kaneko, L. Lavoura, H. Savanaka, M. Tanimoto, {\tt hep--ph/0408123}; and references therein.

\vfill\eject

\end{document}